\documentclass[twocolumn,showpacs,amsmath,amssymb,prl,graphics,graphicx]
{revtex4}
 \usepackage{graphicx}
\usepackage{dcolumn}
\begin{document}
\title{Current in narrow channels of anisotropic superconductors }
\author{V. G. Kogan}
 \affiliation{
    Ames Laboratory - DOE and Department of Physics,   
  Iowa State University, Ames, IA 50011}
\author{V. L. Pokrovsky}
\affiliation{Department of Physics, Texas A\&M University, College Station,
TX 77843-4242\\ and Landau Institute for Theoretical Physics,
Chernogolovka, Moscow Distr. 142432, Russia} 
\begin{abstract}
We argue that in channels cut out of anisotropic single crystal
superconductors and narrow on the scale of London penetration depth, the
persistent current must cause the transverse phase difference provided the
current   does not point in any of the principal crystal
directions. The difference is proportional to the current value and depends
on the anisotropy parameter, on the current direction relative to the
crystal, and on the transverse channel dimension. An experimental set up
to measure the transverse phase is proposed.
\end{abstract}
\pacs{74.20.-x,74.20.De,74.50.+r }
\maketitle
In isotropic superconductors the supercurrent density is proportional to 
the gradient of the gauge invariant phase  $ \nabla \varphi=\nabla \chi
+2\pi {\bf A}/\phi_0$, where $\chi$ is the phase of the order parameter
$\psi = |\psi|e^{i\chi}$, $\phi_0=\pi\hbar c/ |e|$ is the flux
quantum, and ${\bf A}$ is the vector potential: 
\begin{equation}
{\bf j}={2e\hbar\over M}|\psi|^2\,\nabla\varphi\,;
\end{equation}
$M$ is the carrier mass. If the cross-section of a superconducting
channel has small dimensions compared with the London penetration depth
$\lambda$, both ${\bf j}$ and  $\nabla \varphi$ have the only nonzero
components along the channel. For straight channels with the long
dimension along $x$,
$\partial_y\varphi=\partial_z\varphi=0$, i.e., the phase is constant in
the transverse directions.  
   
This, however, is not the case for {\it anisotropic} superconductors,
  where
\begin{equation}
  j_i=2e\hbar\, M_{ik}^{-1}|\psi|^2\,\partial_k\varphi \,. 
\label{e1}
\end{equation}
Here, $M_{ik}$ is the mass tensor, and summation is implied over repeated 
indices. It is convenient to normalize the masses: 
$m_{ik}=M_{ik}/(M_aM_bM_c)^{1/3}=M_{ik}/\,{\overline M} $. Then the
  eigenvalues of
$m_{ik}$ are related  by $m_am_bm_c=1$. In the uniaxial case which we
consider for simplicity, 
$m_a^2m_c=1$; the masses then are expressed in terms of a single parameter, 
the anisotropy ratio $\gamma^2=m_c/m_a$: $m_a=\gamma^{-2/3}$,  
$m_c=\gamma^{4/3}$. 

We invert Eq. (\ref{e1}) to obtain
\begin{equation}
\partial_i\varphi =\frac{{\overline M}}{2e\hbar |\psi|^2}\, m_{ik}\,j_k 
\, ,
\label{e2}
\end{equation}
which shows that the phase gradient  $\nabla \varphi$  and  the current 
${\bf j}$  are not parallel unless both of them point in a  principal
crystal direction.

\begin{figure}[htb]
\includegraphics[angle=0,width=85mm]{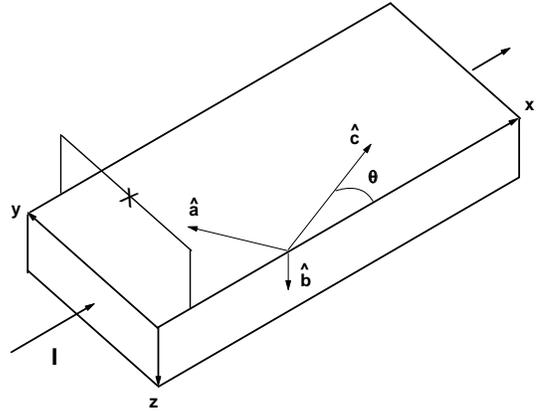}
\caption{\label{f1}The current $I$  in the superconducting channel of
the width $W$  ($0<y<W$) and of the thickness $d$. The crystal axis $ 
b$ coincides with $z$. 
The cross denotes the Josephson junction.}
\end{figure}

Consider a channel of a  rectangular cross-section and
denote by $W$ and $d$ its width in the $y$ direction and the thickness in
the $z$ direction. To further simplify the problem, we take $z$ as one of
principal crystal directions, say $b$. The axes $c$ and $a$ are then
situated in the $xy$ plane as shown in Fig. \ref{f1}. Denoting by $\theta$
the misalignment angle between the $c$ axis and $x$, we readily obtain:
\begin{eqnarray}
m_{xx}&=&
\gamma^{-2/3}(\sin^2\theta+\gamma^2\cos^2\theta)\,, \nonumber \\
m_{yy}&=&
\gamma^{-2/3}(\cos^2\theta+\gamma^2\sin^2\theta) \,, \\
m_{xy}&=&
\gamma^{-2/3}(\gamma^2-1)\sin\theta\cos\theta \,,  \nonumber
\label{e3}
\end{eqnarray}
whereas $m_{zz}=m_b=\gamma^{-2/3}$ and $m_{zx}=m_{zy}=0$.

Let a small supercurrent $I$ be fed into the channel along $x$. Then, Eq.
(\ref{e2}) yields the only non-zero transverse component of $\nabla
\varphi$:
\begin{equation}
\partial_y\varphi =\frac{{\overline M}}{2e\hbar |\psi|^2}\,
m_{yx}\,\frac{I}{Wd}\,.
\label{e4}
\end{equation}
Therefore, the side faces of the channel at $y=0,W$ should possess the phase
difference of
\begin{equation}
\Delta\varphi =\frac{{\overline M}}{2e\hbar |\psi|^2}\,
m_{xy}\,\frac{I}{ d}\, 
\label{e5}
\end{equation}
for any fixed $x$. 

In principle, this phase difference can be recorded by attaching a
superconducting wire  between the points $\{x,0\}$ and $\{x,W\}$. If the
wire carries   the Josephson junction, the current in the wire is 
\begin{equation}
I_w =I_0\sin\Big(\frac{{\overline M\,m_{xy}}}{2e\hbar |\psi|^2d}\,
\, I \Big)\,, 
\label{e6}
\end{equation}
where $I_0$ is the maximum Josephson current. Thus, the current $I_w$ in
the wire oscillates as a function of the driving current $I$ in the channel
with the period
\begin{equation}
\Delta I = \frac{4\pi e\hbar |\psi|^2d }{{\overline M}\,m_{xy}}\,
\, \,. 
\label{e7}
\end{equation}

Near the critical temperature $T_c$, the domain for which the above
formulas are written, the equilibrium order parameter $\psi$ in clean
materials is related to the carrier density $n_e$:
$|\psi|^2\approx n_e\tau$,
$\tau=1-T/T_c\,$. This estimate holds provided $I\ll I_{dp}$ where
$I_{dp}\approx (c\phi_0/16\pi^2\lambda^2\xi)Wd$ is the depairing current. 

To find out conditions under which the current oscillations can be
seen   near $T_c$, we require that both $I$ and $\Delta I$ are small
relative to $I_{dp}$. Taking $W\sim
\lambda(T) \approx \lambda_0/\sqrt{\tau}\approx {\overline M}c^2/4\pi e^2
n_e\sqrt{\tau}$ and $\xi\approx\xi_0/\sqrt{\tau}$, we obtain:
\begin{equation}
 m_{xy} \gg 16\pi/\kappa\,,  
\label{e8}
\end{equation}
 where  $\kappa=\lambda_0/ \xi_0 $. Since $ m_{xy} \sim
m_c=\gamma^{4/3}$, this inequality can be satisfied for strongly
anisotropic materials with large Ginzburg-Landau parameter $\kappa$. 
 
At low $T$'s one can use the London expression for the current
density instead of Eq. (\ref{e1}):
\begin{equation}
     j_i=-\frac{c\phi_0}{4\pi^2\lambda^2}\,m_{ik}^{-1}\Big(\nabla\chi
   +{2\pi\over\phi_0}\,{\bf A}\Big)_k\,,\label{London}
\end{equation}
where $\lambda=(\lambda_a^2\lambda_c)^{1/3}$. We, therefore, can replace
in the above formulas $2e\hbar |\psi|^2/{\overline M}$ with
$c\phi_0/4\pi^2\lambda^2$. In particular, we have 
\begin{equation}
\Delta I  = \frac{c\phi_0 d }{ 2\pi\lambda^2\,m_{xy}}\,
\, \,. 
\label{e10}
\end{equation}
We then obtain a condition  $m_{xy}\gg 8\pi/\kappa$  similar to
(\ref{e8}) in the whole temperature domain; if this condition is not
satisfied, the window of driving currents $I$ for the oscillations of
$I_w$ to occur becomes  narrow or disappears. 

Thus, the period $\Delta I$ is proportional to the
channel thickness $d$ and  to  $\tau$, the temperature
distance from $T_c$. Taking for an estimate realistic values of $\lambda =
2000\,$\AA, $d=1000\,$\AA, and $\gamma=10$, we obtain $\Delta I\approx
1\,$mA for low temperatures; the period shrinks on approaching $T_c$.  

It is worth noting that for the effect to be observable, the wire contacts
must be small relative to $W$. Otherwise the
variation of the phase along the channel will destroy the quantum
coherence. On the other hand, if the point contacts are fixed at different
values of $x$, an additional phase difference 
\begin{equation}
(\Delta\varphi)_x=(\Delta\varphi)_y\,\frac{m_{xx}}{m_{xy}}\,\frac{\Delta
x}{W}
\end{equation}
enters the argument of the sine in Eq. (\ref{e6}); here $\Delta
x=|x_2-x_1|$ and $x_{1,2}$ are the contacts positions. This contribution
changes the period
$\Delta I$ of the Josephson current oscillations, but the very fact of
periodic dependence of $I_w$ on $I$ persists.

If the driving current $I$ changes with time, so does the phase difference
$\Delta\varphi$. Then, the junction in the wire is subject to a voltage 
\begin{equation}
V ={\hbar\over 2e}\partial_t\Delta\varphi= 
\frac{2\pi\lambda^2}{c^2d}\,m_{xy}\,\partial_t I\sim
\frac{2\pi\lambda^2}{c^2d}\,m_{xy}\,\omega I\,, 
\label{e11}
\end{equation}
where $\omega$ is the driving frequency. For the parameters used above
and $\omega\sim 10^9\,$Hz, the voltage may reach values of a few $\mu$V;
this estimate improves if thinner samples and higher temperatures are used:
$V\propto 1/\tau d$.

Writing the driving ac current as $I=I_a \sin\omega t$, we readily obtain:
\begin{eqnarray}
I_w &=&I_0\sin( \varphi_a \sin\omega t) \nonumber\\
&=& 2
I_0\sum_{n=0}^{\infty}   J_{2n+1}( \varphi_a)\, \sin (2n+1)\omega t
\label{e13}
\end{eqnarray}
where $J$'s are the Bessel functions and 
\begin{equation}
\varphi_a= \frac{4\pi^2\lambda^2m_{xy}}{c\phi_0d} \,I_a \,\,. 
\label{e14}
\end{equation}
For small amplitudes $I_a$ of the driving current, one can keep only the
term $n=0$  in the series (\ref{e13}): $I_w = I_0\varphi_a\sin\omega
t$, i.e., the junction response is linear. Since the voltage in this case
is $V=(\phi_0\varphi_a/2\pi c)\omega\cos\omega t$, one can say that the
junction is loaded with an inductance $L=c\phi_0 /2\pi I_0$. This
simple interpretation does not hold if the condition $\varphi_a\ll 1$ is
violated; the junction then should generate odd harmonics.

We have considered a particular case of a misalignment between the
supercurrent direction and the  crystal. Of course, 
situations different from that of Fig. 1 can be readily treated. Our aim  
is to turn the  community attention   to the ``transverse''
phase difference which must accompany supercurrents as long as they are
not parallel to principal crystal directions. We believe that effects
related to the transverse phase in anisotropic materials and their
complexity are  not exhausted by simple examples we describe. In
particular, we did not consider details of the electric field penetration
into the channel (implicitly assuming the penetration length $\ell_E\gg
W$). Given richness of time dependent Josephson
phenomena, one may envisage a host of
  time dependent effects related to the transverse phase. It should also be
stressed that phenomena we describe, happen on the scale of the
London $\lambda$, which might be large as compared to the
coherence length $\xi$, the scale relevant to the well-studied  
  phase slips.\\

We are thankful to A. Larkin for an encouraging discussion and to S. Bud'ko
and U. Welp for the interest in the subject.   The work of V.K. is
supported by the Director of Energy Research, Office of Basic Energy
Sciences, U. S. Department of Energy. The work of V.P. is supported in
part by the NSF Grants DMR 0072115 and DMR 0103455. 

 \end{document}